\documentclass[pre,twocolumn,showpacs,floatfix]{revtex4}

\usepackage{graphicx}

\begin{document}

\title{Some exact results for the velocity of cracks propagating in non-linear
elastic models}

\author{T. M. Guozden}
\author{E. A. Jagla}

\affiliation{Centro At\'omico Bariloche, Comisi\'on Nacional de Energ\'{\i}a 
At\'omica (8400) Bariloche, Argentina}

\begin{abstract}
We analyze a piece-wise linear elastic model 
for the propagation of a crack in a stripe geometry under mode III conditions, in the absence of
dissipation. The model is continuous in the propagation direction and discrete in the
perpendicular direction.
The velocity of the crack is a function of the value of the applied strain. 
We find analytically the value of the propagation velocity 
close to the Griffith threshold, and close to the strain of uniform breakdown. 
Contrary to the case of perfectly harmonic behavior up to the fracture point,  in the piece-wise 
linear elastic model the crack velocity is lower than the sound velocity, reaching this
limiting value at the
strain of uniform breakdown. We complement the analytical 
results with numerical simulations
and find excellent agreement.

\end{abstract}
\maketitle

\section{Introduction}

The velocity of a crack propagating in a brittle material is known to be related to the 
sound velocity in the material. This general statement can be qualitatively justified by
noticing that a crack is a sort of elastic disturbance, although of course of extreme non-linear nature.
Thus it is not surprising that its velocity is related to the velocity of propagation of small
amplitude elastic deformations. However, when we want to be more precise about the relation between
crack velocity and sound velocity, difficulties appear. In text book treatments of linear elastic
fracture mechanics, it is suggested 
that the maximum crack velocity is given by the Raleigh velocity $v_R$.\cite{freund,broberg}
This limit is expected to be achieved at large driving forces (i.e., large applied strain), since
for low driving forces the discrete (atomic) nature of the material may reduce the velocity drastically
(this is called the lattice trapping effect).
Experimentally, an increase of the velocity with the applied strain 
is observed in general, however
the limiting Raleigh velocity is almost never achieved.\cite{fineberg}
Microscopical observation of crack paths in different
kinds of samples reveal one source of this discrepancy: at velocities roughly close to $v_R/3$ 
a straight crack path destabilizes, becoming wandering, and generating side branches at larger velocities.
Some people have claimed \cite{sharon} that if this effect is taken into account (i.e., the true microscopic 
crack path length is larger than the apparent macroscopic length) then the classical prediction is verified.
But this cannot be claimed to be always the case.\cite{kessler1} Even restricting to cracks propagating
is a stationary fashion along a perfectly linear path, a careful analysis reveals that crack propagation velocity cannot 
be determined independently of the microscopic details close to the crack tip.\cite{freund,broberg} 
This means that a purely macroscopic
analysis using continuous approximations for the material leaves the velocity of the crack undetermined.
This is the reason why detailed models of the breaking phenomena at the atomic scale are necessary in determining crack
velocities. 

A class of fully consistent models on which crack velocities can be calculated
(albeit numerically)
are lattice spring models where the material is represented by a set of point masses joined by springs.\cite{slepyan}
These springs can break when some threshold deformation is reached giving rise
to cracks in the form of connected sets of broken springs.

It has been recently established\cite{gao,guozden05} that the propagation velocity in 
this kind of model crucially depends on the presence of anharmonicities 
of the springs. These anharmonicities are also called 
hyperelastic effects. The most spectacular case is that to hyperelastic stiffening (i.e., springs
becoming stiffer at large deformation), that
can produce supersonic crack propagation, something that had been considered
impossible in classical treatments of fracture. However, the case of hyperelastic
softening is by far the expected most common case, since most decohesion
potentials typically interpolate smoothly between the weakly deformed material
and the broken material, in which the elastic constants are formally zero. In this
case, and even in the absence of other effects such as crack velocity oscillation or
crack branching, hyperelastic softening produces a noticeable reduction of
the crack velocity. 

Even in the relatively simple class of lattice spring models, quantitative predictions 
of crack velocity is elusive, since, as already stated, breaking of
the material is a form of non-linear behavior, and it is typically very
difficult to find exact results for non-linear models. 
The situation is even worse in the presence of hyperelasticity,
which is an additional source of non-linear behavior.

In this paper we show that taking a continuous approximation in the propagation direction 
in a class of lattice spring models, some
exact results can be obtained for the crack velocity even in the presence of
hyperelastic softening. These results shed light on the effect of
hyperelasticity on crack propagation, and serve as a starting
point for other (most likely numerical) studies in more realistic models.

We have also implemented the model
numerically and compared the simulated results with the analytical ones, 
finding excellent agreement.

\section{The model}

\begin{figure}[h]
\includegraphics[width=8cm,clip=true]{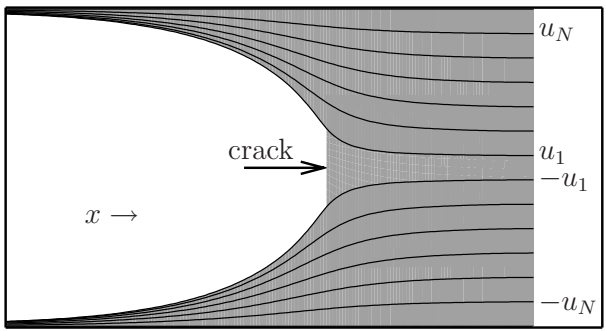}

\caption{A sketch of the model studied: a set of $2N$ continuous 
non-linear elastic chains (represented by the continuous lines) are coupled through perfectly harmonic 
interactions if the vertical distance between chains is lower than a threshold value $u_{bk}$.
If this value is exceeded (this may occur only between chains $u_1$ and $-u_1$) the two chains
decouple defining the crack. In the figure, intact springs are shadowed. The crack 
advances to the right as the system evolves. Boundary conditions 
are fixed displacements imposed at the top and bottom of the figure.}
\label{esquema}
\end{figure}

We consider a lattice spring model in a stripe geometry with fixed displacement mode III boundary conditions.
A continuous description is implemented in the propagating direction (along the
stripe, chosen to be the $x$ direction), whereas a discrete model is considered in the
perpendicular ($y$) direction.
Thus the model consists of a set of a fixed number ($2N$) of continuous elastic 
chains as depicted in Fig. \ref{esquema}. Taking into account
the symmetry of the system, we solve the equations only for the upper half of it, in which each 
chain is labeled by a discrete index $j$, $1\le j\le N$. 
We consider the (out of plane) mode III displacement of the chains,
that is noted $u_j(x)$ for chain $j$. 
Chain $j$ is coupled to the two neighbor chains by harmonic springs. These
springs can break when the coordinate difference between chains is larger than a
breaking threshold that we note $u_{bk}$. We will always assume that the crack propagates in
the middle of the stripe, i.e., between chains $u_1$ and $-u_1$. Our aim is to
find the stable propagation velocity of this crack. Chains $u_N$ and $-u_N$ are
coupled to the lateral sides of the system, on which fixed displacements are applied.
The sides of the system can be formally introduced as chains $u_{N+1}$
(and $-u_{N+1}$), with $u_{N+1}(x)=(N+1/2) \delta$.  
This defines $\delta$ as the nominal
displacement between adjacent chains in the system. Hyperelasticity
comes from the assumption that the spring constant of
a chain changes from a low stretching value $k_0$ when $|du_j(x)/dx|<u_{nl}$, to a
value $k_0\gamma$  when $|du_j(x)/dx|>u_{nl}$.  Thus $u_{nl}$
can be appropriately called the non-linear threshold of the chains. In the present paper we consider only the case 
of hyperelastic softening, namely $\gamma<1$. In short, the model is
defined by the equation:
\begin{eqnarray}
\rho\frac{d^2 u_j(x,t)}{dt^2}&=&\frac{d}{dx}\left[ k_0\eta(|du_j/dx|)
\frac{du_j(x,t)}{dx}\right]+\nonumber\\
&+&\nu(u_{j+1}-u_j)+\nu(u_{j}-u_{j-1}),
\end{eqnarray}
with the functions $\eta$ and $\nu$ defined as
\begin{eqnarray}
\eta(y)= 1~~~~~{\mbox{if}}~~~~~|y|<u_{nl}\nonumber\\
\eta(y)=\gamma~~~~~{\mbox{if}}~~~~~|y|>u_{nl}\nonumber\\
\nu(y)=y~~~~~{\mbox{if}}~~~~~|y|<u_{bk}\nonumber\\
\nu(y)=0~~~~~{\mbox{if}}~~~~~|y|>u_{bk}
\end{eqnarray}
and $\rho$ being the density of each chain.

We want to obtain the solution to this equation when the external strain $\delta$ is in between two
limiting values. The lowest possible value for propagation corresponds to the Griffith's
threshold $\delta_G$, at which the elastic energy available in the system ahead of the crack
equals that stored in the broken springs behind the crack. An easy calculation shows that
$\delta_G=u_{bk}/\sqrt{2N+1}$. On the other hand the maximum external strain that can be supported
by the system is the one that would break the system even in the absence of any pre-existent
crack. Clearly this stress for uniform breaking $\delta_U$ is given by $\delta_U=u_{bk}$.

A few remarks are in order. Our model is obviously anisotropic, as there are
continuous chains along the $x$ direction, whereas the system is discrete
in the $y$ direction. Another source of anisotropy lies in the fact that
hyperelastic softening is introduced only inside the chains, but not in the inter-chain springs. 
We have previously indicated\cite{guozden05} that in fact it is the hyperelasticity in the
propagation direction that drives the non-trivial evolution of the system. There
is no point in introducing hyperelasticity in the interchain springs, as this
has no important effect in the dynamics and complicates greatly the analytical treatment. 
Note also that chains are not allowed to break, it
is only the vertical inter-chain springs that break. This forces the crack to remain in the 
center of the stripe and avoids effects such as crack branching.

The wave velocity inside each chain is given by $V_w\equiv\sqrt{k_0/\rho}$. In the highly stretched
case, the spring constant changes in a factor of $\gamma$, so we can define the stretched
wave velocity $V_w^{\gamma}$ as $V_w^{\gamma}=\sqrt{k_0\gamma/\rho}$.
We will solve the model under the assumption that there is a stable propagation
of a crack in the middle of the stripe, with a velocity $V$. As we will see,
this velocity --if not zero-- will never be larger than $V_w$, nor lower than $V_w^{\gamma}$.

\section{Scaling properties of the solution}

Before presenting the analytical results we have derived, it is interesting
to indicate some constraints on the solution that can be obtained using scaling arguments
only. Let us suppose we have obtained the solution $u_j(x,t)$  corresponding
to a given set of parameters $u_{nl}$, $u_{bk}$, $\gamma$, and some applied
stress $\delta$, and that this solution has a velocity $V$. It is then immediate to verify
that a rescaled solution $\alpha u$ is also a solution of the problem for an applied strain $\alpha \delta$,
with the same velocity $V$ if the parameters $u_{bk}$ and $u_{nl}$ are 
rescaled to $\alpha u_{bk}$, and $\alpha
u_{nl}$. This means that the velocity can be written as a
function of the combinations $u_{bk}/u_{nl}$, and $\delta/u_{nl}$. 

A less
trivial scaling can be obtained by changing a solution of the form $u(x,t)$, to
a new form $w(x,t)\equiv u(Ax,Bt)$, and finding $A$ and $B$ and new coefficients
of the model for $w(x,t)$ to be a solution. The calculation is straightforward, and
we present only the result, that can be stated as the fact that the velocity of
the crack should be of the form
\begin{equation}
\sqrt{1-\left(\frac{V}{V_w}\right)^2}=\sqrt{1-\gamma}f\left (N, \frac{\delta}{u_{nl} \sqrt{1-\gamma}},
\frac{u_{bk}}{u_{nl} \sqrt{1-\gamma}}\right)
\label{scaling}
\end{equation}
where $f$ is an undetermined function of the arguments and where we also made use of the 
previously obtained scaling.

Already from this scaling relation a very important result can be obtained. This corresponds to the case in which
there is no hyperelastic softening, i.e., in which the spring constants within the chains remain always equal to $k_0$.
This can be formally obtained by letting $u_{nl}$ go to infinity. 
Then the r.h.s. of Eq. (\ref{scaling}) becomes
$\sqrt{1-\gamma}f(N,0,0)$. Since obviously in this limit the crack velocity cannot 
depend on $\gamma$, the only possibility is
that $f(N,0,0)=0$, which implies $V=V_w$. This means that in the absence of hyperelastic
effects the crack velocity is equal to the wave velocity for any $\delta>\delta_G$.


\section{Exact Solution for $N=1$}

We present here the exact analytic solution of the previous non-linear model in
the case $N=1$. This will be a reference result for the further discussion of
the more interesting cases with $N>1$.

Assuming a stationary propagation, we introduce the stationary solution $\tilde u(x) \equiv u_1(x-Vt,0)$, 
where $V$ is the
propagation velocity to be determined self-consistently. We choose the new reference system in
such a way  that the crack
tip is located at $x=0$. Thus, we must search for solutions of the piece-wise defined equation
(we eliminate the tilde for simplicity):
\begin{eqnarray}
\left[1-\left(\frac{V}{V_w}\right)^2\right]\frac{d^2u}{dx^2}=&u-\frac{3\delta}{2};~~x<0,~\left|\frac{d u}{dx}\right | <u_{nl}\\
\left[1-\left(\frac{V}{V_w}\right)^2\right]\frac{d^2  u}{dx^2}=&3u-\frac{3\delta}{2};~~x>0,~\left |\frac{d u}{dx}\right | <u_{nl}\\
\left[\gamma-\left(\frac{V}{V_w}\right)^2\right]\frac{d^2  u}{dx^2}=&u-\frac{3\delta}{2};~~x<0,~\left |\frac{d u}{dx}\right | >u_{nl}\\
\left[\gamma-\left(\frac{V}{V_w}\right)^2\right]\frac{d^2 u}{dx^2}=&3u-\frac{3\delta}{2};~~x>0,~\left|\frac{d u}{dx}\right | >u_{nl}
\label{cuatro}
\end{eqnarray}
with the additional condition to be satisfied at the crack tip: $u(0)=u_{bk}/2$.

Non-trivial solutions of this non-linear equation of motion 
can be obtained by matching the solution in the different sectors. 
It can be shown in general that the crack tip (located at $x=0$) must also 
be the point that separates low and high stretching
regions, i.e., $|d u/dx|<u_{nl}$ for $x>0$, and $|d u/dx|>u_{nl}$ for 
$x<0$. In fact, according to the differential equation, when the non-linear threshold
is reached, there is a change of sign in the pre-factor of $d^2 u/dx^2$, that
passes from $[1-(V/V_w)^2]$ to $[\gamma-(V/V_w)^2]$. But $d^2 u/dx^2$ cannot change sign, otherwise the first 
derivative $d u/dx$ would be an extreme at that point, and this is inconsistent since we
assumed $|d u/dx|<u_{nl}$ to the right and $|d u/dx|>u_{nl}$ to the left.
Then the change of sign of the pre-factor of $d^2 u/dx^2$ must be compensated by a change of sign
on the right hand side of the equation, and this is only possible at the point where
the system is breaking and the right hand side changes from $3u-3\delta/2$ to 
$u-3\delta/2$. This justifies our statement that exactly at the crack tip, the
value of $|d u/dx|$ is $u_{nl}$.

For $x>0$ the solution of the differential equation has the form
\begin{equation}
u(x)=\frac {\delta}{2}-\left(\frac{\delta}{2} -\frac{u_{bk}}{2}\right ) 
\exp{\left [-\frac{x\sqrt{3}}{\sqrt{1-(V/V_w)^2}}\right ]},
\label{uderecha}
\end{equation} 
where we have already used the constraint $u(0)=u_{bk}/2$, 
and from the requirement $|d u/dx|_{x=0}=u_{nl}$ we get the velocity as
\begin{equation}
V(\delta)=V_w\sqrt{1-\frac 34\left(\frac{u_{bk}-\delta}{u_{nl}}\right)^2}
\label{v}
\end{equation} 
Note that the velocity is independent of the value of $\gamma$, and that it is 
consistent with Eq. (\ref{scaling}), as it must be. We also see explicitly that $V=V_w$ if 
$u_{nl}\rightarrow \infty$.
To check that this is a consistent solution, we must verify that there is a reasonable
form of $u(x)$ for $x<0$. 



When $\delta$ is large enough,
the solution for 
$x<0$ consists of a concatenation of similar pieces of the form
\begin{eqnarray}
u(x)&=&\frac{3\delta}{2}+ \frac{u_{bk}-3\delta}{2}\cos\left (\frac{x}{\sqrt{(V/V_w)^2-\gamma}}\right
)-\nonumber\\
&-&u_{nl}\sqrt{(V/V_w)^2-\gamma}\sin\left(\frac{x}{\sqrt{(V/V_w)^2-\gamma}}\right)
\label{triang}
\end{eqnarray} 
\begin{figure}[h]
\includegraphics[width=8cm,clip=true]{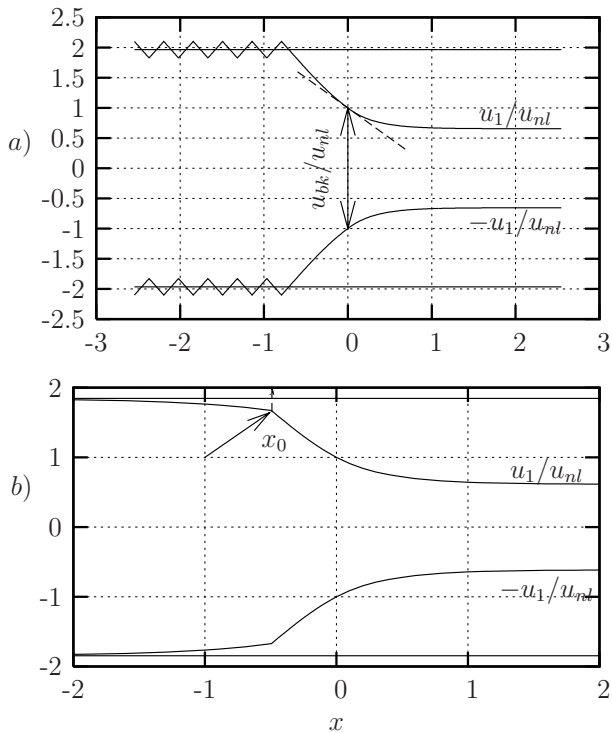}
\caption{Exact solutions for $N=1$, $u_{bk}/u_{nl}=2$ and  $\delta/u_{nl}=1.31$ in $a)$ and $\delta/ u_{nl}=1.23$ in $b)$. 
Jumps  of $\frac{du_1}{dx}$ in $a)$ satisfy a momentum conservation condition given by Eq. (\ref{jump1}).
In $b)$, the jump  in $\frac{du_1}{dx}$ 
at $x=x_0$ is described by Eq. (\ref{discont}). In this case, only the portion of chains between 
$x=x_0$ and $x=0$ is in the hyperelastic regime. In both cases $\left.\frac{du_1}{dx}\right|_{x=0} =-u_{nl}$ and $u_1(0)=\frac{u_{bk}}{2}$.}
\label{sing}
\label{serru}
\end{figure}
This kind of solution is sketched in Fig. \ref{serru}.
When two pieces of this form are matched together, the derivative
of $u$ has a jump. A momentum conservation condition must be satisfied at those points. In fact, 
the integral of the force on an infinitesimal piece of chain during the time in which it passes through 
the singular point, must be equal to the change of momentum. It is obtained that 
this conservation requires in fact that the derivative of $u$ a jump. The value of
the derivative is equal on both sides, and its value is
\begin{equation}
|du/dx|=u_{nl}(1-\gamma)/((V/V_w)^2-\gamma).
\label{jump1}
\end{equation}
Note in Fig. \ref{serru}$(a)$ how the sort of triangular oscillation extends to $x\rightarrow -\infty$, i.e., 
the excess of elastic energy in the system remains in the
form of kinetic and elastic energy far behind the crack tip.

When $\delta$ is reduced, the amplitude of the oscillation described by Eqs. (\ref{triang})
and (\ref{jump1}) reduces also, and it vanishes at a particular value of $\delta$.
For $\delta$  values lower than this, the previous solution is not valid. 
It turns out that in this case
the solution for $x<0$ is singular in our continuous
system. A way to ``regularize" this singular solution is to 
consider the  continuous chain as the
limit of a discrete chain of point masses. When the number of masses per unit length 
goes to infinity, we can describe the solution that is obtained in the following way [see Fig. \ref{sing}(b)]:
there is a region for $x_0<x<0$
in which the solution has the previous form given in Eq. (\ref{triang}).
For $x<x_0$, the solution returns to the linear regime, and is composed by a smooth part and a singular part. 
The singular part
is an oscillation that has an amplitude which goes to zero in the continuum limit, but a
frequency that diverges in this limit, in such a way that it can carry (in the form of
kinetic energy) the excess of elastic 
energy in the system. The regular part has the form
\begin{equation}
u(x)=\frac {3\delta}{2}-\left(\frac{3\delta}{2} -\frac{u_{bk}}{2}\right ) 
\exp{\left [\frac{x\sqrt{2}}{\sqrt{(V/V_w)^2-\gamma}}\right ]},
\label{uregular}
\end{equation} 
and the matching conditions at $x_0$ for this regular part correspond to have continuity of the function, i.e, 
$u(x_0^-)=u(x_0^+)$, and a jump in the derivative given by
\begin{equation}
\left.\frac{du}{dx}\right|_{x_0^-}[1-(V/V_w)^2]-\left.\frac{du}{dx}\right|_{x_0^+}[\gamma-(V/V_w)^2]=
u_{nl}(1-\gamma)
\label{discont}
\end{equation}
which is obtained using the same kind of conservation arguments that led to Eq. (\ref{jump1}).
In Section VI we will 
show results of numerical simulations that confirm and clarify further this behavior.

We see in the solution for $x<0$ (Eq. \ref{triang}) that when $V\rightarrow V_w^\gamma$ the
frequency of the oscillation for $x<0$ diverges. In fact, if according to Eq. \ref{v}
the velocity would be lower than $V_w^\gamma$, then 
neither the solutions  given by Eq. (\ref{triang}) nor Eq. (\ref{uregular}) exist.
It can be shown that in this case the velocity of crack
propagation is actually $V=V_w^{\gamma}$.
This regime is not particularly interesting to us, and from
now on we will always assume to have chosen values of $\gamma$ such that $V>V_w^{\gamma}$.

Note that according to Eq. (\ref{v}) the crack velocity at the 
Griffith's threshold $\delta_G=u_{bk}/\sqrt{3}$ is finite if
$u_{bk}/u_{nl}<2/(\sqrt 3-1)$, and is given by 
\begin{equation}
V(\delta_G)=V_w \sqrt{1-\left(\frac{u_{bk}}{u_{nl}}\right)^2\frac{(\sqrt 3 -1)^2}{4}}
\label{vdg} 
\end{equation}
 This is in contrast to cases in which the system 
is discrete in the direction of crack propagation. In those cases, due to 
lattice trapping effects the crack cannot propagate
if the Griffith's threshold is not overpassed by a finite amount.

The values in Eq. (\ref{v}) for the velocity as a function of $\delta$ must 
be compared with the result that is obtained in the absence of hyperelastic effect, namely 
$V(\delta)=V_w$. 
The non-trivial result contained in Eq. (\ref{v}) is a consequence of
hyperelasticity in the system. We will see that the same qualitative effects exist in the
more interesting cases with $N>1$.

\section{Exact results for $N>1$}

The previous case $N=1$ is a good starting point in which the analytical solution can be
worked out in full detail. But obviously, if we are interested in modeling a macroscopic
system we should study the case of a large number of chains. 

\begin{figure}[h]
\includegraphics[width=8cm,clip=true]{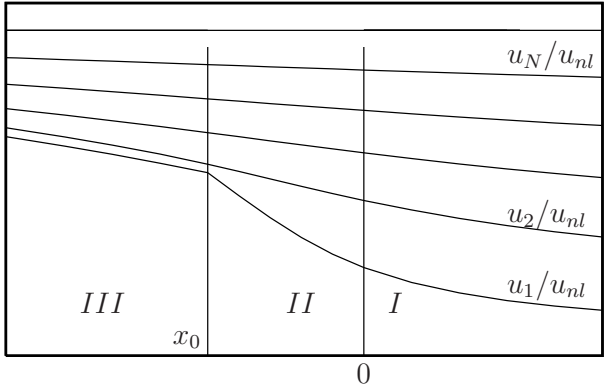}
\caption{A case with $N=5$ and $\delta$ close to $\delta_G$.
For clarity, we only plot the upper half of chains, as the others are symmetric.
Only the  chain adjacent to the crack, and only in region II, explores the nonlinear regime. 
As $\delta\rightarrow \delta_G$, $x_0 \rightarrow 0$.
}
\label{zonas}
\end{figure}

\begin{figure}[h]
   \includegraphics[width=8cm,clip=true]{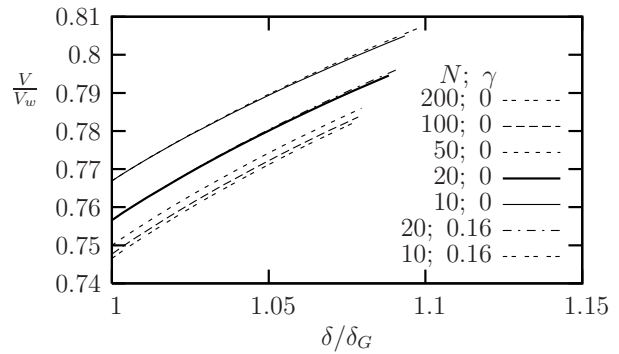}
   \caption{Analytical results for the velocity $vs$ $\delta / \delta_G$,
   for $u_{bk}/u_{nl}=4/3$.
   It is seen that the dependence
   on $\gamma$ is very weak.
   Through all the range in which we plot the data
   only chain $u_1$ (and $-u_1$) explores the hyperelastic regime.
   }
   \label{resAndg}
\end{figure}

\begin{figure}[h]
   \includegraphics[width=8cm,clip=true]{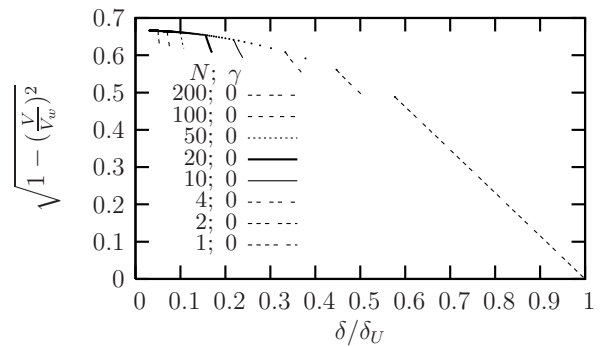}
   \caption{Same as in the previous figure, plotted as a function of $\delta/\delta_U$. 
   }
   \label{resAn}
\end{figure}

For $N>1$ the exact value of the velocity for arbitrary $\delta$ cannot be obtained, in
general. 
However, we can provide exact results in some neighborhood of $\delta_G$ and 
$\delta_U$. Consider first the case $\delta \sim \delta_G$. 
Let us concentrate on one half (the upper one) of the system, since the other is symmetric.
Sufficiently close to the Griffith's threshold, only one chain (the one adjacent to the
crack) will enter the hyperelastic region. In this regime the problem can be separated
in three sectors as shown in Fig \ref{zonas}.
 We match the solutions, requiring continuity of the
function and derivative of $u_j(x)$, except for the derivative of $u_{1}(x)$, in which
(as in the $N=1$ case) a discontinuity of the derivative of the form (\ref{discont}) exists between regions II and III. 
The solution obtained will be valid
as long as no chain other than the first enters the non-linear regime and $u_{1}(x)<u_{2}(x)$.
The width of zone II is determined as part of the solution.
The problem stands as a system of $4N+1$ 
nonlinear algebraic equations, which we solve to any desired accuracy through 
an iterative method. The results for the velocity are
plotted in 
Fig. \ref{resAndg} as a function of $\delta/\delta_G$, and in
Fig. \ref{resAn} as a function of $\delta/\delta_U$.
We plot data in the full range in which the method is reliable and gives the exact value of the velocity.
As it can be observed, the velocity is only weakly dependent on
$\gamma$ (always assuming $V_w^{\gamma}<V$).

It is interesting to observe from Fig \ref{resAndg}, 
that even in the limit $N\rightarrow\infty$ our method provides the solution 
in a finite range of $\delta$.
This means that in all this interval, for a system of
infinite chains, there is a single one that explores the hyperelastic regime, and is responsible for the full reduction of the
velocity from $V_w$ to the actual value.


\begin{figure}[h]
   \includegraphics[width=8cm,clip=true]{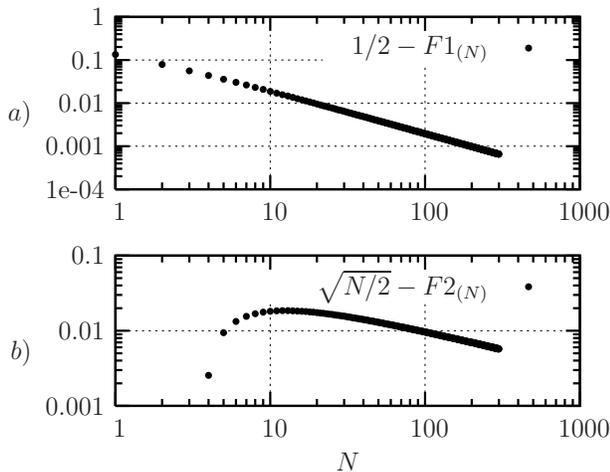}
   \caption{ Plot of the functions $F_1(N)$ and $F_2(N)$ [see Eq2. (\ref{eqF1}) and (\ref{eqF2})].
   }
   \label{F1}
\end{figure}

In the limit $\delta\rightarrow \delta_G$, we have $x_0\rightarrow 0$, and region II shrinks to zero.
In this limit we obtain the exact values of the velocity and its derivative
with respect to $\delta$ by solving a linear system of $2N$ algebraic equations.
Both $V(\delta_G)$ and $\left. \frac
{dV}{d\delta}\right| _{\delta_G}$
turn out to be independent of $\gamma$. 
From this independence and the scaling form Eq. (\ref{scaling}), we can conclude that 
\begin{equation}
V(\delta_G,N)=V_w\sqrt{1-\left( \frac{u_{bk}}{u_{nl}}F_1(N) \right)^2}
\label{eqF1}
\end{equation}
and
\begin{equation}
\left.\frac{ d\sqrt{1-(V/V_w)^2}}{d\delta}\right|_{\delta_G,N}=\frac{F_2(N)} {u_{nl}}
\label{eqF2}
\end{equation}

The values of the functions $F_1$ and $F_2$ for different $N$ are shown in Fig. \ref{F1}.
The limiting value $F_1(N\rightarrow\infty)=1/2$ can be obtained analytically through an appropriate
analysis of the equations in this limit. Extrapolation of the finite $N$ exact values of $F_2(N)$
suggests also that $\lim_{N\rightarrow\infty}\left(F_2(N)/\sqrt{N/2}\right)=1$, but we have not verified it analytically.

We can then write:
\begin{eqnarray}
\left.\sqrt{1-\left(V/V_w\right)^2}\right|_{\delta_G, N\rightarrow \infty}= \frac{u_{bk}}{2 u_{nl}}\label{limites1}
\\
\left.\frac{d\sqrt{1-(V/V_w)^2}}{d(\delta/\delta_G)}\right|_{\delta_G,N\rightarrow \infty}= \frac{u_{bk}}{2u_{nl}}
\label{limites2}
\end{eqnarray}
We emphasize again the reduction of the
velocity from the value $V_w$ due to the hyperelastic effect. This effect disappears if $u_{nl}\rightarrow\infty$.

\begin{figure}[h]
\includegraphics[width=8cm,clip=true]{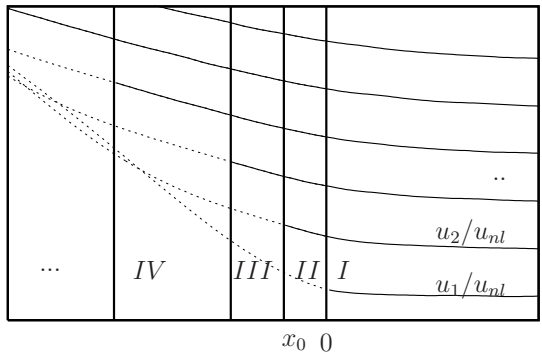}
\caption{Sketch of the configuration for $\delta$ close to $\delta_U$. 
In sectors I, II, III, etc., we have 0, 1, 2, etc. chains in the hyperelastic regime
(dotted lines).
The velocity for $\delta$ close to $\delta_U$ can be obtained analyzing sectors I 
and II only, as explained in the text.
}
\label{F_lim_ubk}
\end{figure}

A second limiting case can be solved analytically, and that is the asymptotic form of the velocity very close to
$\delta_U$. 
The analysis is based again in 
matching the solutions of the piece-wise linear equation. 
The situation is sketched in Fig \ref{F_lim_ubk}.
In sector I ($x>0$) all chains are in the linear regime. In sectors II, III, etc., chains enter 
the nonlinear regime sequentially, starting from the one adjacent to the crack. 

When $\delta\rightarrow\delta_U$ (and $V/V_w \rightarrow 1 $), it can be seen that 
the solution of the equations for sector II are $N-1$ exponential modes with a 
diverging decaying constant, 
and a trigonometric mode with finite frequency. Taking into account that 
the width of region II  (namely $|x_0|$) remains finite even for $V\rightarrow V_w$, 
we can neglect exponential modes that grow toward negative $x$.
This allows us to obtain the velocity in this limit by solving a system of $2N$
linear equations, matching the solutions between regions I and II only.


The result we obtain is that to lowest order in $\delta-\delta_U$, the velocity can be written as
\begin{equation}
\left.\sqrt{1-\left(\frac{V}{V_w}\right)^2}\right|_{\delta\rightarrow\delta_U}=
\frac{\delta-\delta_U}{u_{nl}} F_3(N)  
\label{atubk}
\end{equation}
\begin{figure}[h]
\includegraphics[width=8cm,clip=true]{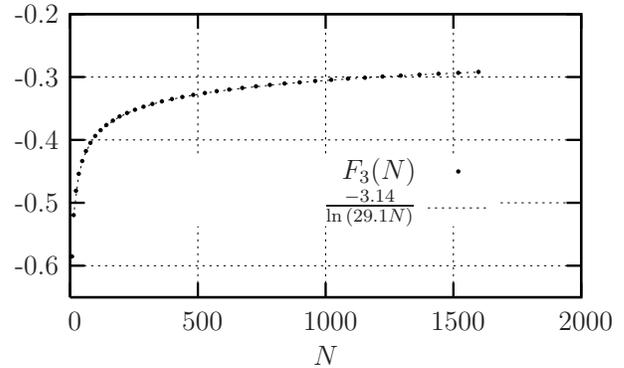}
\caption{Plot of $F_3(N)$ (see Eq. \ref{atubk}). The data follow a logarithmic trend, vanishing for $N\rightarrow\infty$ as $1/\ln N$
as the fitting (dotted line) shows.}
\label{F3}
\end{figure}
where $F_3$ is another $N$-dependent dimensionless function.
Note that this result is again independent of $\gamma$, and consistent with the general expression in Eq. (\ref{scaling}).
Values of  $F_3(N)$ are plotted in Fig \ref{F3}.
By analyzing in more detail the $N\rightarrow\infty$  limit, it can be shown that $F_3$ goes to zero as $1/\ln(N)$.

\section{Comparison with Numerical Results}

Although the main results of our work are the analytical findings of the previous
section, we include here some results of numerical simulations for two reasons:
first of all, some of the results of the previous section are not fully intuitive, and
then we think it is clarifying to check them against a numerical simulation. Secondly,
numerical simulations can be done in the full range between $\delta=\delta_G$ and
$\delta=\delta_U$, filling the gap between the two analytical limits. 

\begin{figure}[h]
   \includegraphics[width=8cm,clip=true]{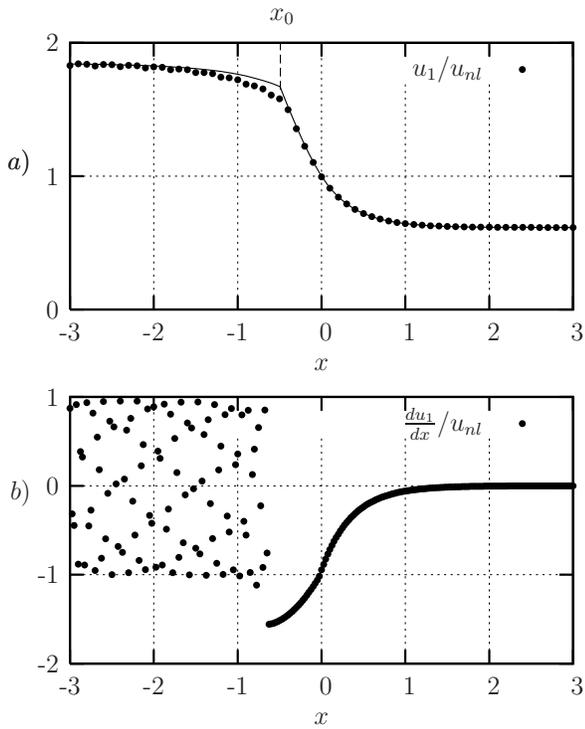}
   \caption{$a)$: Analytical (line) and numerical results (points) for $N=1$,
    $\delta/u_{nl}=1.23$ and $u_{bk}/u_{nl}=2$ ($\kappa=1/1600$ in the numerics).
   For clarity, only one every four points of the
   simulated system is shown.
   $b)$ The plot of $du/dx$, from the simulations.
   We calculate the derivative in the discrete system for chain $j$ as
   $du/dx\equiv \sqrt{\kappa}\left[u_{i+1,j}-u_{i,j}\right]$, where the subindex $i$ is
   the discretization along the $x$ axis. Note that 
   the only piece of chain in the non-linear regime is that with $x_0<x<0$. 
   The oscillation for $x<x_0$
   carries (in the form of kinetic energy) the excess of elastic energy that is
   present in the system.
   }
   \label{singII}
\end{figure}
\begin{figure}[h]
   \includegraphics[width=8cm,clip=true]{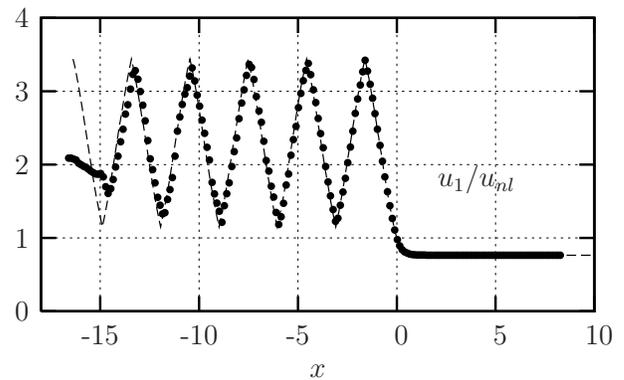}
   \caption{$a)$: Analytical (line) and  numerical results (points) for $N=1$ with
    $\delta/u_{nl}=1.53$ and $u_{bk}/u_{nl}=2$ 
   ($\kappa=1/1600$ in the numerics). For clarity, only one every five points of the
   simulated system is shown.
   The solution behaves as described in Eq. (\ref{triang}).
   The numerical results differ at the left border of the system, 
   because the system is finite in the numerical simulation.
   }
      \label{serruII}
\end{figure}
\begin{figure}[h]
\includegraphics[width=8cm,clip=true]{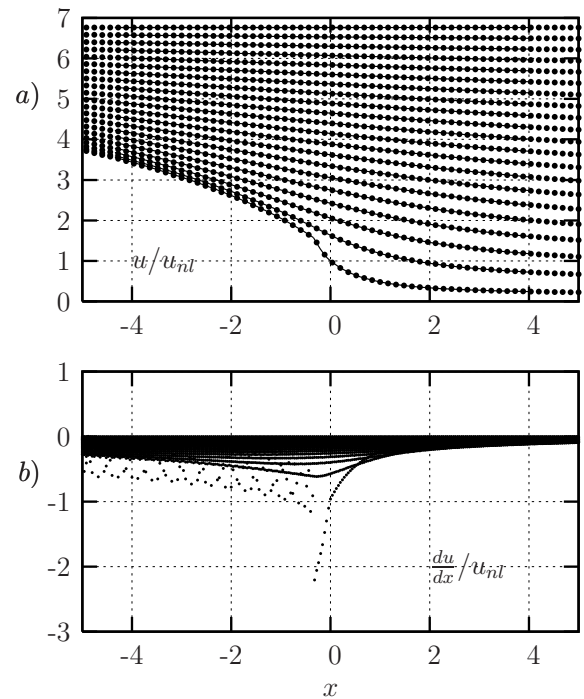}
\caption{$a)$: Analytical (lines) and numerical results (points) for $N=20$, $\delta/u_{nl}=0.33$  and $u_{bk}/u_{nl}=2$ 
($\kappa=1/625$ in the numerics). For clarity, only one every four points of the simulated system is shown.
At this strain value only the first chain enters the non-linear regime.
$b)$ Numerical values of $\frac{du}{dx}$,
   where the singular oscillation behind the crack 
   is observable.
}
\label{c20}
\end{figure}

In our numerical simulations we are forced to consider a system that is discrete also in
the $x$ direction. We do this by introducing softer springs in the $x$
direction than in the $y$ direction. The ratio between horizontal and vertical spring
constants will be noted $\kappa$, and it is a measure of the degree of anisotropy of the lattice.
For $\kappa=1$ the lattice is isotropic, whereas for $\kappa\rightarrow 0$ we recover the
continuous limit of the analytical treatments. As we will see, keeping a finite but small
$\kappa$ is also an appropriate form of regularizing the singular results that may appear in the
continuum case.

In Figs. \ref{singII} and \ref{serruII} we present results for $N=1$.
 They  compare very well with the analytical results of the previous Section.
Note in particular in Fig. \ref{singII}(b) the oscillation of $du/dx$ for $x<x_0$. This represents the singular
behavior of the analytical solution that we have discussed previously. This oscillation is 
seen in the plot of $du/dx$, but is hardly visible in the plot of $u(x)$ itself, as its amplitude 
goes to zero with $\kappa$. 
In the case of Fig. \ref{serruII} we see how the abrupt change on the derivative of the analytical solution is
very well reproduced in the numerical simulations, supporting the prescription given by Eq. (\ref{jump1}). 
In Fig. \ref{c20} we show superimposed analytical and numerical results for $N=20$, for a value of $\delta$ in which a single chain explores the non-linear regime.
Again the agreement is very good. Note also in this case the oscillation 
in $du/dx$ for the first chain behind the crack. 

\begin{figure}[h]
\includegraphics[width=8cm,clip=true]{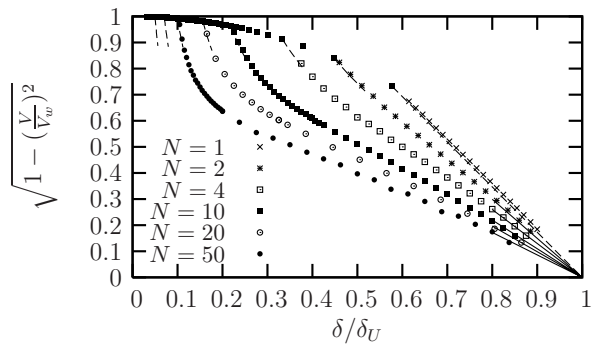}
\caption{Numerical results for different number of stripes with $\gamma=0$ and $u_{bk}/u_{nl}=2$.
Analytical results close to $\delta_G$
are plotted as dashed lines, whereas the analytical asymptotic slopes at $\delta_U$ [Eq.(\ref{atubk})] 
are shown by continuous lines. 
The agreement is seen to be excellent, considering the discreteness of the numerical model.
Note that for the parameters chosen the velocity at $\delta_G$ goes to zero when $N\rightarrow \infty$ [Eq. (\ref{limites1})].
}
\label{resNum}
\end{figure}

The numerical results for crack velocities are plotted 
on top of the analytical results in Fig. \ref{resNum}. We see that they 
agree very well with the analytical values, providing a link between 
the $\delta\sim\delta_G$ and $\delta\sim \delta_U$ regions.

\section {The continuous limit: $N\rightarrow\infty$}

From the finite $N$ results of the previous sections 
we can try to obtain the behavior of a `macroscopic' material by studying the limit $N\rightarrow\infty$.
We must keep in mind however that the results we are about to discuss still depend strongly 
on the microscopic details of the model, other microscopic realization giving rise probably to different 
macroscopic behavior. In fact,
we have already emphasized that the fracture of a macroscopic object cannot be described completely 
in terms of a continuum description, since microscopic details of the breaking process at 
the crack tip are always relevant. 
In any case, taking the $N\rightarrow\infty$ limit in our model provides us with one possible realization
of a continuum system that we want to analyze now.

The most important quantity we can analyze in the large $N$ limit is the dependence of the crack velocity
on the normalized strain  $\delta/\delta_G$.
This is in fact a directly accessible experimental quantity.
We already have at hand some analytical results about this (see Eqs. \ref{limites1} 
and \ref{limites2}). We know the value
of $V$ at $\delta_G$ (which is strictly lower than $V_w$), and that of $dV/d(\delta/\delta_G)$ at 
$\delta=\delta_G$ (which is finite). We also know that eventually $V$ reaches the value $V_w$ for sufficiently large 
$\delta/\delta_G$. But the extremely slow decay with $N$ of 
$dV/d\delta$ at $\delta=\delta_U$ (Eq. (\ref{atubk})) allows to infer that $V(\delta/\delta_G)$ will reach the value $V_w$ also 
very slowly. We present here a non-rigorous argument, which we think reproduces the right tendency.
First of all note from the numerical results of Fig. \ref{resNum} that the asymptotic value of $V$ close to $\delta_U$
is a reasonable upper bound for the velocity, i.e. [see Eq. (\ref{atubk})],
\begin{equation}
\left.\sqrt{1-\left(\frac{V}{V_w}\right)^2}\right|_{\delta/\delta_G, N}
\gtrsim \frac{u_{bk}}{u_{nl}}\left(\frac{\delta/\delta_G}{\sqrt{2N+1}}-1\right)F_3(N)
\label{envol}
\end{equation}
\begin{figure}[h]
\includegraphics[width=8cm,clip=true]{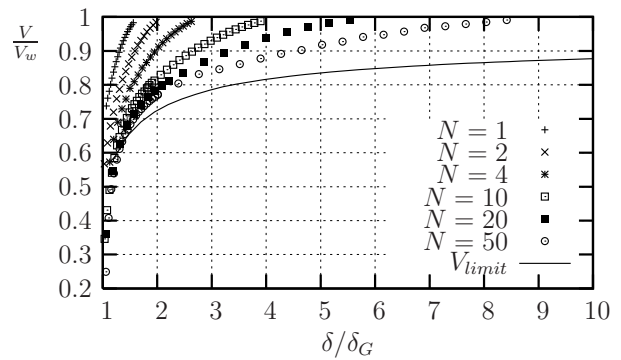}
\caption{Same numerical results as in Fig. \ref{resNum}, for different number of stripes with $\gamma=0$ and
$u_{bk}/u_{nl}=2$,
but plotted as a function of $\delta/\delta_G$.
The numerical results suggest that, given $\delta/\delta_G$, the velocity is a decreasing function of $N$.
This evidence allows us to plot an upper bound (continuous line) for the results in the $N\rightarrow\infty$ case (see the text for details).
}
\label{logN}
\end{figure}
On the other hand, when plotted as a function of $\delta/\delta_G$ (Fig. \ref{logN}), the numerical
results suggest that, for a fixed value of  $\delta/\delta_G$, the velocity is a decreasing function of $N$. 
We think this is rigorous result, although we do not have a proof of it. Accepting this statement as valid,
we can obtain an upper bound for the velocity in the $N\rightarrow\infty$ limit by maximizing the right hand side of
Eq. (\ref{envol}) with respect
to $N$ for each value of $\delta/\delta_G$. The result is plotted in Fig. \ref{logN} as a continuous line. 
For large $\delta/\delta_G$, the leading analytical form can be obtained as
\begin{equation}
\left .V(\delta/\delta_G,N)\right|_{N\rightarrow\infty, \delta/\delta_G\rightarrow\infty}\sim
V_w\left[1-\alpha\left(\frac{u_{bk}}{u_{nl}}\frac{1}{\ln{\delta/\delta_G}}\right)^2\right]
\end{equation}
where $\alpha$ is a numerical constant. This
is in fact an extremely slow convergence to the limiting value $V_w$. 
We think this is a very important result. It tells that strictly speaking, hyperelastic softening does not reduce the
limiting value of the velocity for sufficiently large applied strain. However, in view of its extremely slow convergence
to this limit, from a practical point of view we can say that hyperelastic softening produces an appreciable reduction 
of the limiting velocity. In
particular, we see that this reduction of the velocity does not exist 
when hyperelasticity is absent (namely, for $u_{nl}\rightarrow\infty$).

\section{Discussion and Conclusions}

We have analyzed the effect of hyperelastic softening in a model of crack propagation in a 
stripe geometry under mode III fixed displacement boundary conditions. 
The model is continuous in the propagation direction
and has a finite number of chains in the perpendicular direction.
The two central chains of the stripe can decouple when they separate more than a
critical distance $u_{bk}$, generating a crack in the model.

In the case in which the chains are perfectly harmonic
the velocity of crack propagation is equal to the wave velocity $V_w$ 
in the full range of external strain $\delta$ between the Griffith's
threshold $\delta_G$ and the strain of uniform breakdown $\delta_U$.

We have studied how this result is affected by 
the inclusion of hyperelastic softening in the chains, namely, a 
softening of the spring constant of the chains when the stretching
is greater than a threshold value. We have provided analytical results 
in some cases, and complemented them with numerical simulations.

For the case of a single chain the full analytical solution has been 
worked out. It is clearly seen already in this simple case that 
hyperelastic softening reduces the velocity from the harmonic case. 
Now the velocity has a non-trivial dependence on $\delta$, and 
becomes equal to $V_w$ only at $\delta_U$.

We have given the analytical solution for the velocity 
in neighborhoods of $\delta=\delta_G$ and $\delta=\delta_U$. The main results in this case 
are the following. The crack velocity at $\delta_G$ is strictly lower than $V_w$. It decreases as a 
function of the number of chains but may well be finite in the 
$N\rightarrow\infty$ limit for some range of the parameters of 
the model. There is a finite range of $\delta/\delta_G$ 
 in which only the chain adjacent 
to the crack enters the hyperelastic regime. 
This range remains finite for large $N$.
This means that our analytical treatment provides the
exact value of the velocity in a finite range around $\delta/\delta_G=1$ even in the $N\rightarrow\infty$ case.

The crack velocity tends always to $V_w$ when $\delta\rightarrow\delta_U$. 
In the large $N$ limit this can be stated as the fact that $V$ tends to $V_w$
for $\delta/\delta_G\rightarrow\infty$. However, ours estimations show that this
convergence is very slow, namely like $\sim 1/\ln^2(\delta/\delta_G)$.

\section{Acknowledgments}

The authors acknowledge the  financial support of CONICET (Argentina).

\end{document}